\documentstyle[11pt,aaspp4]{article}
\def\gsim{\;\rlap{\lower 2.5pt
 \hbox{$\sim$}}\raise 1.5pt\hbox{$>$}\;}
\def\lsim{\;\rlap{\lower 2.5pt
   \hbox{$\sim$}}\raise 1.5pt\hbox{$<$}\;}
\newcommand\beq{\begin{equation}}
\newcommand\eeq{\end{equation}}
\def\lya{Ly$\alpha$~}

\begin{document}

\title{Observational Signatures of the First Quasars}
\author{Zolt\'an Haiman\altaffilmark{1} and Abraham Loeb\altaffilmark{2}}
\medskip
\affil{Astronomy Department, Harvard University, 60 Garden Street,
Cambridge, MA 02138}
\altaffiltext{1}{email:zhaiman@cfa.harvard.edu}
\altaffiltext{2}{email:aloeb@cfa.harvard.edu}

\begin{abstract}

We study the observational signatures of a potential population of
low-luminosity quasars at high redshifts in a $\Lambda$CDM cosmology.
We derive the evolution of the quasar luminosity function at fainter
luminosities and higher redshifts than currently detected, based on
three assumptions: (1) the formation of dark--matter halos follows the
Press--Schechter theory, (2) the ratio of central black hole mass to
halo mass is the same for all halos, and (3) the light--curve of
quasars, in Eddington units, is universal. We show that a universal
light--curve provides an excellent fit to the observed quasar
luminosity function at redshifts $2.6<z<4.5$.  By extrapolating the
evolution of this luminosity function to higher redshifts
($4.5<z<20$), we find that the associated early population of
low-luminosity quasars reionizes the universe at a redshift $z\sim12$.
The reprocessing of the UV light of these quasars by dust from early
type II supernovae, distorts the microwave background spectrum by a
Compton $y$--parameter, $y\sim 10^{-5}$, comparable to the upper limit
set by COBE.  The Next Generation Space Telescope could detect tens of
quasars from redshifts $z>10$ per square arcminute, with its proposed
1nJy sensitivity at 1--3.5$\mu$m.  Absorption spectra of several such
quasars would reveal the reionization history of the universe.

\end{abstract}

\keywords{cosmology: theory -- quasars: general}

\centerline{Submitted to {\em The Astrophysical Journal}}

\section{Introduction}

One of the outstanding problems in cosmology is the nature of the
first generation of astrophysical objects which appeared when the
universe first transformed from its initial smooth state to its
current clumpy state.  Although we have observational data on bright
quasars and galaxies out to redshifts $z\sim5$ (Schneider, Schmidt \&
Gunn 1991; Franx~et~al.~1997), and on the linear density fluctuations
at redshift $z\sim10^3$ (Bennett~et~al.~1996), there is currently no
direct evidence as to when and how the first structures formed, and
what kind of objects were responsible for end of the ``dark age'' of
the universe (Rees~1996).

Popular Cold Dark Matter (CDM) models for structure formation predict
the appearance of the first baryonic objects with masses
$M\sim10^5{\rm M_\odot}$ at redshifts as high as $z\sim30$; objects
with progressively higher masses assemble later (cf.  Haiman, Thoul \&
Loeb, and references therein). Following virialization, the gas in
these objects can only continue to collapse and fragment if it can
cool on a timescale shorter than the Hubble time.  In the metal--poor
primordial gas, the only coolants that satisfy this requirement are
neutral atomic hydrogen (H) and molecular hydrogen (${\rm
H_2}$). However, ${\rm H_2}$ molecules are fragile, and are easily
photodissociated throughout the universe by trace amounts of starlight
(Haiman, Rees, \& Loeb 1996, 1997; Gnedin \& Ostriker 1997).  Hence,
most of the first stars are expected to form inside objects with
virial temperatures $T_{\rm vir}\gsim10^4$K.  Depending on the details
of their cooling and angular momentum transport, the gas in these
objects is expected to either fragment into stars, or form a central
black hole exhibiting quasar activity. Conversion of a small fraction
($\sim 1-10\%$) of the gas into stars or quasars could reionize the
universe, and strongly affect the entropy of the intergalactic medium.

In two previous papers (Haiman \& Loeb 1997a, hereafter HL97; and Loeb
\& Haiman, 1997, hereafter LH97), we explored the impact of the first
stars on the reionization history of the universe. In this paper, we
study the related signatures of early quasars, and compare their
effects to those expected from the stars. For both types of sources,
we calibrate the total amount of light they produce based on data from
redshifts $z\la 5$. The efficiency of early star formation is
calibrated based on the observed metallicity of the intergalactic
medium (Songaila \& Cowie 1996, Tytler~et~al.~1995), while the early
quasars are constrained to match the quasar luminosity function at
redshifts $z\la 5$ (Pei 1995).  In HL97 and LH97 we have studied a
range of standard Cold Dark Matter (CDM) cosmologies, but in this
paper we focus on a particular cosmological model with a cosmological
constant, namely the ``concordance model'' of Ostriker \& Steinhardt
(1995).  Within this model, we predict the number of high--redshift
low-luminosity quasars using the Press--Schechter formalism (Press \&
Schechter 1974).

We assume that the luminosity history of each black hole depends only
on its mass, and postulate the existence of a universal quasar
light--curve in Eddington units. This approach is motivated by the
fact that for a sufficiently high fueling rate, quasars are likely to
shine at their maximum possible luminosity, which is some constant
fraction of the Eddington limit, for a time which is dictated by their
final mass and radiative efficiency. If the radiative efficiency were
determined by a universal accretion geometry, then the bright phase of
quasars would admit such a universal light--curve.  We also assume
that the final black hole mass is a fixed fraction of the total halo
mass, and allow this fraction to be a free parameter.  Based on this
minimal set of assumptions, we demonstrate that there exists a
universal light curve, for which the Press--Schechter theory provides
an excellent fit to the observed evolution of the luminosity function
of bright quasars between redshifts $2.6<z<4.5$.  Furthermore, our
fitting procedure results in black hole to halo mass ratios close to
the typical values found in local galaxies (Kormendy et al. 1997;
Magorrian et al. 1997). Given this ratio and the fitted light--curve,
we then extrapolate the observed LF to higher redshifts and low
luminosities.

Previous work on modeling the evolution of the quasar LF has been
mostly phenomenological, and involved fitting ad--hoc parametric
functions to the observations (Pei 1995, Boyle~et~al.~1991).  Small \&
Blandford (1992) had used a universal quasar light--curve, together
with the observed LF evolution, to infer the birth--rate of quasar
black holes.  Haehnelt \& Rees (1993; see also Efstathiou \& Rees
1988) adopted the Press--Schechter theory for halo formation,
specified the quasar light--curve a priori, and introduced a redshift
and halo--mass dependence to the black hole formation efficiency, in
order to fit the observed evolution of the quasar LF.  The postulated
(ad--hoc) exponential decrease in the efficiency for black hole
formation in small halos (with circular velocities less than 400~km/s)
could be attributed to the expulsion of gas from shallow potential
wells by supernova-driven winds or mergers.  However, this
prescription appears to be violated in nearby low-mass galaxies, such
as the compact ellipticals M32 and NGC 4486B.  In particular, van der
Marel et al.  (1997) infer a black hole mass of $\sim 3.4\times
10^6{\rm M_\odot}$ in M32, which is a fraction $\sim 8\times 10^{-3}$
of the stellar mass of the galaxy, $\sim 4\times 10^8{\rm M_\odot}$,
for a central mass-to-light ratio of $\gamma_V=2$; while Kormendy et
al.  (1997) infer a black hole mass of $6\times 10^8{\rm M_\odot}$ in
NGC 4486B which is a fraction $\sim 9\%$ of the stellar mass.  The
efficiency of black hole formation in these systems had been
comparable if not higher than in galaxies with deeper potential wells.
Observations of galactic nuclei in the local universe imply black hole
masses which are typically a fraction $\sim 6\times 10^{-3}$ of the
total baryonic mass of their host (see summary of dynamical estimates
in Fig. 5b of Magorrian et al. 1997; and the maximum ratio between a
quasar luminosity and its host mass found by McLeod 1997).  We
therefore adopt the simplest possible assumption, namely that the
black hole to halo mass ratio is constant, and {\it infer} the quasar
light--curve from observations rather than postulate it.  The actual
formation process of low--luminosity quasars was addressed by
Eisenstein \& Loeb (1995), and Loeb (1997).

Based on the extrapolated quasar luminosity function, we quantify the
effects of the first generation of low-luminosity quasars.  In
particular, the UV radiation produced by these quasars above the Lyman
limit can lead to the reionization of the intergalactic medium (IGM),
and to a number of observable signatures (Carr, Bond \& Arnett 1984).
First, the resulting optical depth of the universe to electron
scattering at $z\la 20$, $\tau_{\rm es}$, damps the microwave
background anisotropies on scales $\la 10^\circ$ (Efstathiou \& Bond
1987; Kamionkowski, Spergel, \& Sugiyama 1994; Hu \& White 1996).  The
recently detected hint for a Doppler peak in ground--based microwave
anisotropy experiments can already be used to set the constraint
$\tau_{\rm es}\la 1$ (Scott, Silk \& White 1995; Bond 1995).  Future
satellite experiments (such as MAP or Planck\footnote{See the
homepages for these experiments at http://map.gsfc.nasa.gov and at
http://astro.estec.esa.nl/SA-general/Projects/Cobras/cobras.html.})
will be able to probe values of $\tau_{\rm es}$ as small as a few
percent if polarization data is gathered (Zaldarriaga, Spergel, \&
Seljak 1997).  The quasar UV flux is also absorbed by intergalactic
dust, and subsequent thermal emission by this dust at longer
wavelengths could introduce a substantial spectral distortion to the
cosmic microwave background (CMB) radiation.  Such a distortion due to
population III stars have been investigated extensively in the past
(Wright 1981; Adams et al. 1989; Bond, Carr \& Hogan 1991; Wright et
al. 1994; LH97), and its magnitude is currently constrained by COBE to
have a Compton $y$--parameter $<1.5\times10^{-5}$.  Finally, direct
detection of high--redshift low--luminosity quasars will be feasible
with the Next Generation Space Telescope, whose proposed sensitivity
in the 1--3.5$\mu$m range is $\sim$1nJy.

The purpose of this paper is to quantify the above signatures of
high-redshift low-luminosity quasars, and to compare them with those of
early stars. The paper is organized as follows.  In \S~2, we describe the
procedure used to find the universal quasar light--curve, and the resulting
fits to the observed quasar LF at $z\la 5$.  In \S~3, we calculate the
reionization history of the intergalactic medium due to early quasars, and
compare it to stellar reionization.  In \S~4, we discuss the effects of the
early quasars on the CMB, including the smoothing of its anisotropies and
the distortion of its spectrum.  In \S~5, we predict the number counts of
high--redshift quasars down to the sensitivity limit of the Next Generation
Space Telescope.  Finally, \S~6 summarizes the main conclusions and
implications of this work.

\section{Quasar Light--Curve}

Our goal is to calculate the quasar luminosity function at faint
magnitudes, ($\log (L_{\rm B}/{\rm L_{B,\odot}})\lsim 11.5$) and at
high redshifts ($z\gsim4.5$), outside the regime of current
observations. For this purpose, we introduce a new model which
extrapolates the observed LF of bright quasars at redshifts
$2.6<z<4.5$, based on the Press--Schechter theory and the assumption
that all quasars admit the same universal light--curve in Eddington
units.

The formation rate of dark--matter halos is given by the
Press--Schechter formalism.  We assume that black hole formation is
restricted to halos whose virial temperature exceeds $T_{\rm vir}\gsim
10^4$K, in which atomic and Bremsstrahlung cooling allows the gas to
sink to the center of the potential well.  This virial temperature
threshold corresponds to a minimum halo mass $M_{\rm halo}\gsim M_{\rm
min}=10^{8}{\rm M_\odot}[(1+z)/10]^{-3/2}$.  We further assume for
simplicity that the ratio of final black hole mass to halo mass,
$\epsilon\equiv M_{\rm bh}/M_{\rm halo}$ is the same for all halos.
Estimates of black hole masses in nearby galaxies
(Kormendy~et~al.~1997; Magorrian et al.  1997) imply that massive
black holes weigh roughly a fixed fraction of their bulge mass, over a
wide range of bulge masses $10^9{\rm M_\odot}\la M_{\rm bulge}\la
2\times 10^{12}{\rm M_\odot}$. Since bulges are the oldest stellar
components of galaxies, it is reasonable to assume that a fixed
fraction of the baryonic content of early galaxies ends up in the
central black hole.  Note, however, that this assumption differs from
the model presented by Haehnelt \& Rees (1993), in which $\epsilon$ is
assumed to decline exponentially for low-mass galaxies.

The change in the comoving number density of black holes with masses
between $M_{\rm bh}$ and $M_{\rm bh}+dM_{\rm bh}$, between redshifts
$z$ and $z+dz$, is governed by the derivative,
\beq
\left.\frac{d^2 N_{\rm bh}(M_{\rm bh},z)}{dM_{\rm bh}dz}\right|_{M_{\rm bh}}=
\left\{\matrix{
{1\over \epsilon} 
\left.\frac{d}{dz}
\frac{dN_{\rm ps}(M,z)}{dM}\right|_{M=\epsilon^{-1} M_{\rm
bh}}
\hfill&({\rm for}~M_{\rm bh}\ge \epsilon M_{\rm min})\hfill\cr 
0
\hfill&({\rm for}~M_{\rm bh}< \epsilon M_{\rm min})\hfill,\cr}\right.
\label{eq:rate}
\eeq where $dN_{\rm ps}/dM$ is the Press-Schechter mass function.  The
actual halo formation rate is larger than the derivative ${d\over
dz}(dN_{\rm ps}/dM)$, since this derivative includes a negative
contribution from merging halos.  However, at high redshifts collapsed
objects are rare, and the merger probability is low. We have compared
the above expression with the more accurate result for the halo
formation rate given by Sasaki (1994), and confirmed that the
difference in the rates is negligible for the high redshifts and halo
masses under consideration here.  We therefore use
equation~(\ref{eq:rate}) to describe the black hole formation rate at
high redshifts.

We define $\phi(L,z)dL$ to be the number of quasars per unit comoving
volume at redshift $z$, with intrinsic luminosities between $L$ and
$L+dL$.  The luminosity function, $\phi(L,z)$, is linked to the black
hole formation rate through the light--curves of the quasars.  For
simplicity, we make the minimal assumption that all quasars admit the
same light--curve in Eddington units, i.e. their luminosity is
proportional to their final mass (see \S 1).  Under this assumption,
the luminosity $L(t)$ of a quasar with a black hole mass $M_{\rm bh}$
at a time $t$ after its birth can be written as
\beq 
L(t)=M_{\rm bh}f(t)= \epsilon M_{\rm
halo}f(t),~~~~~~~~~~~~~~~~~~~~~~(M_{\rm halo}>M_{\rm min}) 
\label{eq:lightcurve}
\eeq where $f(t)$ is a function of time.
This leads to the expression for the LF, 
\beq
\phi(L,z)=\int_{z}^{\infty} \int_{\epsilon M_{\rm min}}^{\infty} 
dM_{\rm bh} dz^{\prime} 
\frac{d^2N_{\rm bh}}{dM_{\rm bh}dz^{\prime}}\delta[L-M_{\rm bh} 
f(t_{z,z^{\prime}})],
\eeq
where $t_{z,z^{\prime}}$ is the time elapsed between the redshifts $z$ and
$z^{\prime}$. By integrating over $z^\prime$, we obtain
\beq
\phi(L,z)=\int_{\epsilon M_{\rm min}}^{\infty} 
\frac{dM_{\rm bh}}{M_{\rm bh}\left|\dot{f}\right|_{t_\star}} 
\left.\frac{d^2N_{\rm bh}}{dM_{\rm bh}dz}\right|_{z_\star} 
\left.\frac{dz}{dt}\right|_{z_\star},
\label{eq:conv1}
\eeq 
or, equivalently, by integrating over $M_{\rm bh}$ we get,
\beq
\phi(L,z)=\int_{0}^{t(z)}
\frac{dt}{f(t)}\left.\frac{dz}{dt}\right|_t
\left.\frac{d^2N_{\rm bh}}{dM_{\rm bh}dz}\right|_{M_{\rm bh}=\frac{L}{f(t)}},
\label{eq:conv2}
\eeq where $\dot{f}=df/dt$, $t_\star$ is defined by the condition
$M_{\rm bh} f(t_\star)\equiv L$, and $z_\star$ is defined through the
relation $t_\star\equiv t(z)-t(z_\star)$.  We found it useful to
utilize both of the above expressions for $\phi(L,z)$, since a
comparison between them can be used to check for the numerical
accuracy of the calculation.  The only unknown quantities are
$\dot{f}$ in equation~(\ref{eq:conv1}), and $f$ in
equation~(\ref{eq:conv2}). Therefore, either of these two equations
can be solved to find the $f(t)$ that would fit best the observational
data.  Since the equations are implicit, we have found their solution
iteratively.  We have optimized our initial guess for $f(t)$ as
follows.

The function $d^2 N_{\rm bh}/dM_{\rm bh}dz$ peaks at a mass scale $M_{\rm
pk}=\epsilon M_{\rm nl}(z)$, where $M_{\rm nl}(z)$ is the non--linear
mass--scale at the corresponding redshift.  Assuming that this function is
sharply peaked around $M_{\rm pk}$, i.e. substituting $d^2 N_{\rm
bh}/dM_{\rm bh}dz~\propto \delta(M_{\rm pk}[z])$ into
equation~(\ref{eq:conv1}), yields
\beq \phi(L,z)\propto\frac{1}{M_{\rm
pk}\left|\dot{f}\right|}\propto\frac{1}{\left|dL/dt\right|}, 
\eeq
or 
\beq t\propto\int_{L}^{\infty}\phi(L^{\prime},z)dL^{\prime}.
\label{eq:f0}
\eeq 
In the last step, we assumed that $dL/dt<0$, i.e. that the quasar's
luminosity decreases monotonically with time.  Equation~(\ref{eq:f0})
yields the time $t$ it takes the quasar luminosity to drop
down to $L=M_{\rm pk}f$, and provides a ``first guess'' solution for $t(f)$
and hence for $f(t)$.  The physical interpretation of the LF in this case
is that it merely reflects the time spent by the quasars in each luminosity
interval.

\begin{figure}[b]
\vspace{2.6cm}
\includegraphics{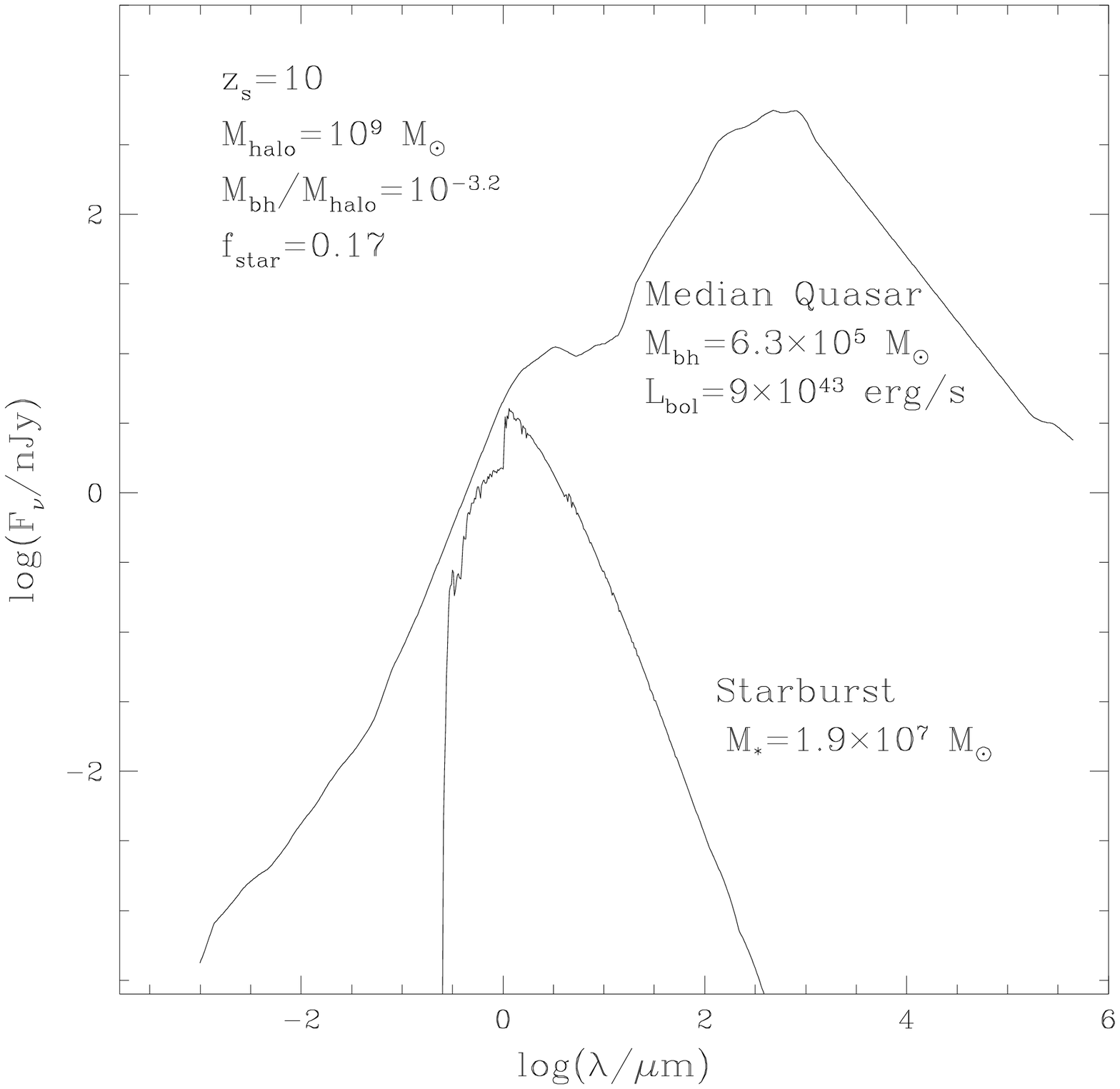}
\vspace*{4.5in}
\caption[Spectra] {\label{fig:spectrum} Comparison of the observed
spectral fluxes from a typical quasar and a starburst in a source located
at redshift $z_s=10$.  The star--formation efficiency, $f_{\rm star}=0.17$,
was fixed so as to produce an average of 1\% solar metallicity in the IGM
at $z=3$. The black hole formation efficiency, $M_{\rm bh}/M_{\rm
halo}=10^{-3.2}$, was found by our fitting procedure to the observed
quasar luminosity function (see text for details).}
\end{figure}

To find the light--curve $f(t)$, we need to specify $\phi(L,z)$.
Given the observational data, the inferred luminosity function depends
on the assumed cosmology and quasar spectrum (the latter due to
$k$-corrections).  In this paper, we adopt the ``concordance model''
of cosmological parameters highlighted by Ostriker \& Steinhardt
(1995), i.e.  a flat $\Lambda$CDM model with a slightly tilted power
spectrum ($\Omega_{\rm m},\Omega_\Lambda, \Omega_{\rm
b},h,\sigma_{8h^{-1}},n$)=(0.35, 0.65, 0.04, 0.65, 0.87, 0.96).
Convenient expressions in this model for the differential volume
element, luminosity distance, and time--redshift relation were given
in terms of elliptic integrals by Eisenstein (1997).  An accurate
arithmetic fitting formula for the growth function that appears in the
Press--Schechter formula, is given by Carroll~et~al.~(1992).  For the
average quasar spectrum, we have used the ``median'' spectrum derived
from observed spectra of 47 quasars by Elvis~et~al.~(1994). This
spectrum is shown in Figure~\ref{fig:spectrum}.

Pei (1995) provides a fitting formula for the B--band (0.44 $\mu$m
rest--wavelength) luminosity function of observed quasars,
$\phi(L_{\rm B},z)$, including the redshift evolution both for a flat
and an open $\Omega_{\rm tot}=0.2$ cosmology.  For conversion to
Eddington units, we note that a 1${\rm M_\odot}$ black hole, shining
at Eddington luminosity with the median flux distribution of
Elvis~et~al.~(1994), has a B--band luminosity of $5.7\times10^3$ times
the solar B-band luminosity.  Pei (1995) assumes power--law spectra
for the quasars, with a slope $\alpha$=0.5 in the flat model, and
$\alpha$=1.0 in the open model.  Since near the B--band the
Elvis~et~al.~(1994) template spectrum is very close to an $\alpha$=1
power--law, we used the open--universe fitting formulas from Pei
(1995), with a rescaling of the luminosities and volume elements for
our $\Lambda$CDM cosmology.  Note that the exact conversion of the
intrinsic LF to a different cosmology would require a
spectrum--dependent $k$-correction, as well as rescaling the
luminosity and the volume element, for the redshift of each individual
quasar.  However, we have found that using the mean redshift in each
redshift bin, and assuming the $\alpha$=1.0 power--law for our quasar
spectrum, provided the correct LF to within several percent.

\begin{figure}[b]
\vspace{2.6cm}
\includegraphics{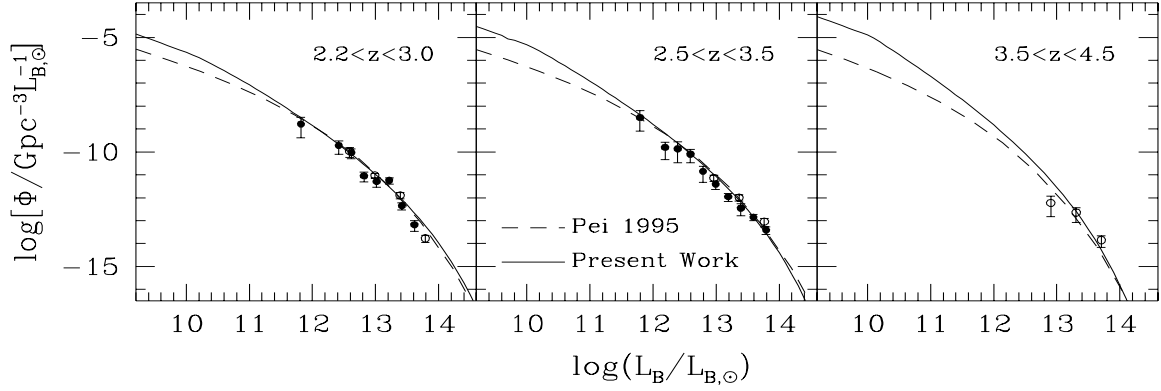}
\vspace*{4.5in}
\caption[LF] {\label{fig:peicomp} The observed and extrapolated
evolution of the quasar luminosity function in a $\Lambda$CDM
cosmology.  The data is taken from Pei (1995), and the dashed lines
show the parametric fitting function from this reference.  The solid
lines show our fits based on the Press--Schechter formalism
and the universal light--curve shown in Fig.~3.}
\end{figure}

\begin{figure}[b]
\vspace{2.6cm}
\includegraphics{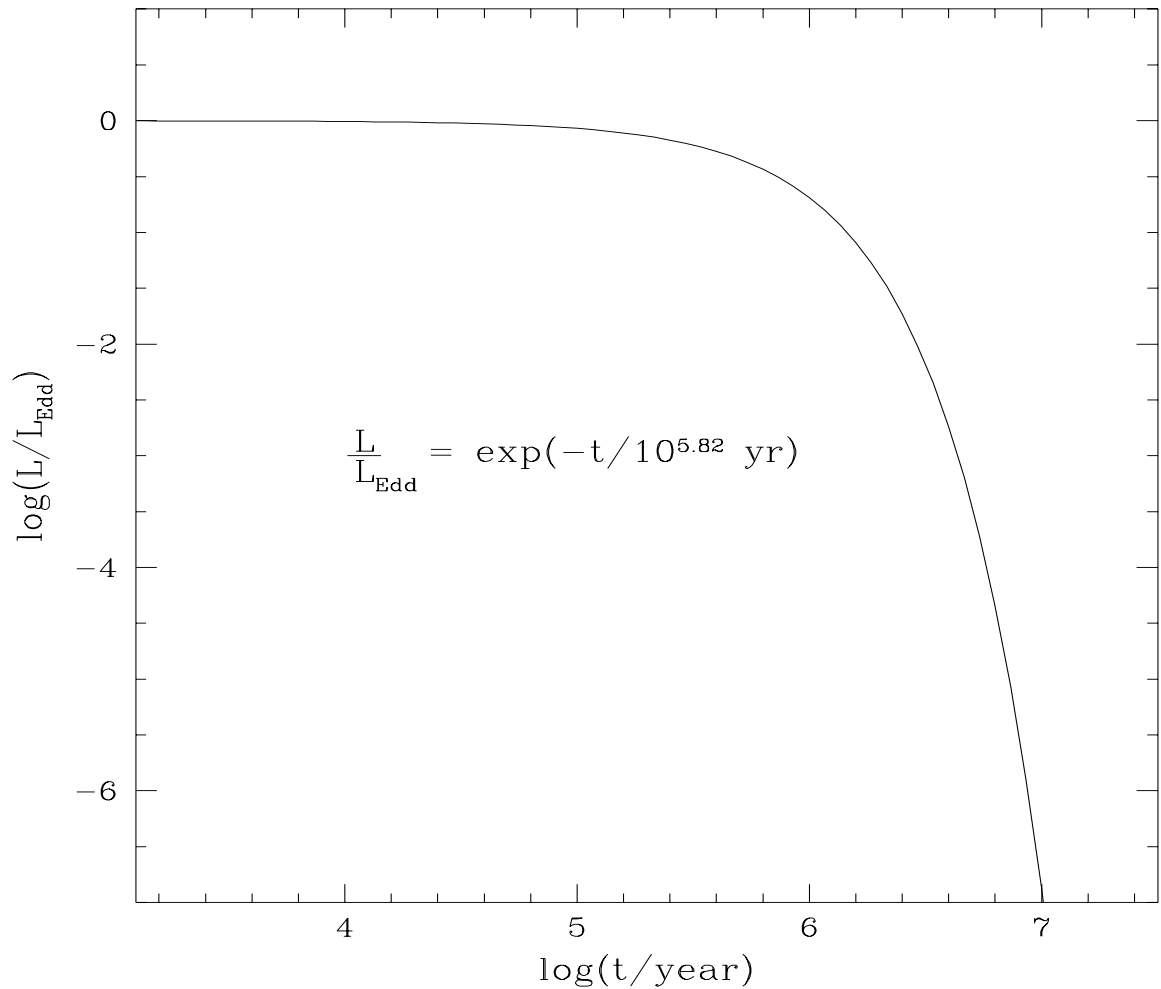}
\vspace*{4.5in}
\caption[Light Curve] {\label{fig:lcurve} The best--fit quasar
light--curve in Eddington units, that results in the luminosity functions
shown by the solid lines in Figure~\ref{fig:peicomp}.  The Eddington time
was not used to rescale the time axis, because it does not depend on the
black hole mass, and is a constant for a universal radiative efficiency.}
\end{figure}

Figure~\ref{fig:peicomp} shows the resulting fitting formulas for
$\phi(L_{\rm B},z)$, together with the observational data points for
the LF, in the $\Lambda$CDM cosmology at redshifts $z$=2.6, 3, and 4.
Using the fitting formula derived by Pei (1995),
equation~(\ref{eq:f0}) yields an initial guess for $f(t)$.  The
numerical solution for this guess can be fitted by a parameterized
function, and substituted back into equation~(\ref{eq:conv1})
or~(\ref{eq:conv2}). The parameters of the fit can then be iteratively
adjusted so as to optimize the match between the resulting $\phi(L,z)$
and the observed quasar LF at $z=2.6, 3$, and $4$.  Using this
procedure, we have found that an excellent fit is achieved through a
simple single--parameter function for the light--curve,
\beq f(t) =
{L_{\rm Edd}\over M_{\rm bh}} \exp\left(-{t\over t_0}\right),
\label{eq:ffit}
\eeq where $L_{\rm Edd}\equiv 1.4\times 10^{38}~{\rm erg~s^{-1}}
(M_{\rm bh}/M_\odot)$ is the Eddington luminosity for the
final\footnote{During the initial growth of the black hole, its mass
and hence its Eddington limit, are smaller. This is not in conflict
with our prescribed light--curve, as long as most of the light is
emitted during a late phase of duration $t_0$, during which the black
hole shines close to the Eddington limit of its final mass, $M_{\rm
bh}$. The inferred low value of $t_0$ implies that much of the black
hole mass is accumulated via accretion with a low radiative efficiency
(e.g., an advection-dominated accretion flow, cf. Narayan 1996 and
references therein).} black hole mass $M_{\rm bh}$.  Our model,
therefore, has only two free parameters, $t_0$ and $\epsilon$.  The
best fit values of these parameters are $t_0=10^{5.82}$ yr and
$\epsilon=10^{-3.2}$. The best--fit light--curve is shown in
Figure~\ref{fig:lcurve}, while the resulting luminosity functions at
$z=2.6$, 3, and 4, are shown by the solid lines in
Figure~\ref{fig:peicomp}.

Our procedure results in fits to the LF comparable in quality to the
original ad--hoc parametric fits given by Pei (1995).  It is rather
remarkable that, in the range of observed luminosities, the two curves
are almost indistinguishable, even though at fainter magnitudes our
curves deviate from the Pei (1995) fits, and predict many more quasars
than the extrapolation of those fitting formulae.  Another gratifying
feature of our approach is that -- with the underlying assumption that
the flat portion of the light--curve corresponds to the Eddington
luminosity limit of the final black hole -- it results in a prediction
for the black hole to halo mass, $M_{\rm bh}/M_{\rm halo}=10^{-3.2}$,
or black hole to gas mass $M_{\rm bh}/M_{\rm gas}=10^{-3.2}\Omega_{\rm
m}/\Omega_{\rm b}=5.4\times10^{-3}$. This is close to the value
derived from observational data, $6\times10^{-3}$, for the bulges of
nearby galaxies (Kormendy~et~al.~1997; Magorrian et al. 1997), which
might have formed at about the same time as quasars.  There are,
however, two shortcomings of our model. First, since it is based on
the Press--Schechter formalism, it predicts that the number of bright
quasars (those with masses above $M_{\rm pk}$, or $L_{\rm
B}\gsim10^{13}{\rm L_{B,\odot}}$) is steadily increasing down to
$z=0$.  Hence, by construction, it cannot explain the decrease in the
number density of bright quasars at low redshifts ($z\lsim2$). This
decrease might be associated with the decline in merger rates at low
redshifts (Carlberg 1990); or it might also be a result of the
expulsion of cold gas from galaxies through supernovae-driven winds
due to the peak in the star formation activity around $z\sim 2$ (Madau
1997).  However, since our predictions focus on high redshifts
($z\ga5$), we ignore this shortcoming of the model.  Second, the
best--fit light--curve predicts a quasar lifetime of only
$\sim6.6\times10^5$ years, about two orders of magnitude shorter than
the Eddington time at the usually assumed value of the radiative
efficiency, 10\%.  The average radiative efficiency over the growth
history of the black hole, is $\sim$0.1\%, two orders of magnitude
smaller than the value for thin disks (Frank~et~al.~1992). This may
imply that quasar black holes accumulate most of their mass in an
accretion flow with a low radiative efficiency, such as advection
dominated flows (Narayan 1996, and references therein).  Although
$t_0$ is an unusually short lifetime for the bright phase of quasars,
it is not in conflict with existing observations.  Furthermore,
increasing the lifetime of quasars would enhance all of the effects
that we describe below.  In this sense, we are being conservative by
adopting our best--fit light--curve.

The evolution predicted by equation~(\ref{eq:conv2}) for the quasar LF
between redshifts $2.6<z<18$, is shown in Figure~\ref{fig:lfs}.  The
break at $L_{\rm B}\sim10^{10}{\rm L_{B,\odot}}$ is introduced by our
imposed low--mass cutoff $M_{\rm halo}\ge 10^{8}{\rm
M_{\odot}}[(1+z)/10]^{-3/2}$.  In Figure~\ref{fig:nz}, we show the
evolution of the number density of quasars brighter than absolute blue
magnitude $M_{\rm B}=-27.5$ and $-22.5$ ($L_{\rm B}=10^{13.2}{\rm
L_{B,\odot}}$ and $10^{11.2}{\rm L_{B,\odot}}$, respectively).  The
solid lines in this figure show the number densities predicted by our
model, while the dotted lines show the corresponding number densities
based on Pei's (1995) fitting formulas.  This figure demonstrates that
although our model agrees with the observational data between
$2.6<z<4.5$, it predicts a substantially larger abundance of quasars
at higher redshifts and fainter magnitudes, than the formal
extrapolations of the fitting formulae given by Pei (1995).

\section{Reionization}

In this section, we derive the reionization history of the IGM in the
presence of early quasars.  The method of calculation follows closely
the one introduced in HL97 for the case of stellar reionization.  We
refer the reader to this reference for more details.

Each ionizing source creates an expanding Str\"omgren sphere around
itself.  Reionization is complete when the separate HII regions
overlap (Arons \& Wingert 1972).  Since the neutral fraction inside
each HII region is small ($\sim10^{-6}$) and the mass-weighted
clumpiness of the IGM is still small at $z\ga 10$ (Gnedin \& Ostriker
1997), the average ionized fraction at any redshift is simply given by
the filling factor of the HII regions.  The reionization history
depends on the formation rate of quasars, the light--curve of ionizing
photons from quasars, and the recombination rate in the IGM, which are
all functions of redshift.

The production rate of ionizing photons per quasar follows from the
quasar spectrum and light--curve shown in Figures~\ref{fig:spectrum}
and~\ref{fig:lcurve}.  The production rate per solar mass of the black
hole, $dN_{E>13.6{\rm eV}}/dt=6.6 \times 10^{47}~M_{\rm bh}~f(t)$
photons ${\rm M_\odot^{-1}~s^{-1}}$, is shown as the solid line in
Figure~\ref{fig:nion}.  For comparison, we also show the analogous
rate per stellar mass from the composite stellar spectrum in our model
of a zero--metallicity starburst.  Note that although the quasar rate
is initially higher than the stellar rate by two orders of magnitude,
it drops much more rapidly at later times.  The total number of
ionizing photons produced by a quasar is $\sim1.3\times10^{61}~M_{\rm
bh}/{\rm M_\odot}$, only an order of magnitude higher than that of
stars, $\sim1.3\times10^{60}~M_{\rm stars}/{\rm M_\odot}$.

We assume that recombinations take place only in the homogeneous
IGM. This includes the implicit assumption that all of the ionizing
photons escape from the neighborhood of the quasar.  While this would
be a poor assumption for stellar radiation, it is justified for the
much more intense (and harder) quasar emission.  A simple estimate can
be used to show that the quasar host would be ionized shortly after
the quasar turns on. Indeed, the known quasars do not show any sign of
associated (galactic) HI absorption in their spectrum, while stellar
spectra of high redshift galaxies often do.  An additional assumption
in obtaining the recombination rate is that the IGM is homogeneous,
i.e. the clumping factor for ionized hydrogen is unity, $C_{\rm
HII}\equiv\left<n_{\rm HII}^2\right>/\left<n_{\rm HII}\right>^2=1$.
This is justified in numerical simulations, which show that $C_{\rm
HII}$ rises significantly above unity only when the collapsed fraction
of baryons is large, i.e. at $z\la 10$ (see Fig.~2 in Gnedin \&
Ostriker 1997)

The evolution of a cosmological ionization front for a time--dependent
source is governed by the equation,
\beq n_{\rm H}\left(\frac{dr_{\rm i}}{dt}-H(t)r_{\rm i}\right)=
\frac{1}{4\pi r_{\rm i}^2}\left(\frac{dN_{\gamma}}{dt}(t)-
\frac{4}{3}\pi r_{\rm i}^3 \alpha_{\rm B}n_{\rm H}^2 \right),
\label{eq:front}
\eeq where $n_{\rm H}$ is the neutral hydrogen number density in the
IGM, $r_{\rm i}$ is the physical (i.e. not comoving) radius of the
ionization front, $H(t)$ is the Hubble expansion rate at cosmic time
$t$, $dN_{\gamma}/dt$ is the ionizing photon rate, and $\alpha_{\rm
B}=2.6\times10^{-13}~{\rm cm^3~s^{-1}}$ is the recombination
coefficient of neutral hydrogen at $T=10^4$K (see HL97, or Shapiro \&
Giroux 1987 for more details).  The solution of this equation yields
the ionized volume per source, $4/3\pi r_{\rm i}^3$. A summation over
all sources provides the total ionized volume within the IGM.

The results for the stellar case depend on the assumed star formation
efficiency, which we normalize based on the observed metallicity of the IGM
at $z\sim 3$ (HL97).  In individual Ly$\alpha$ clouds with $N_{\rm
HI}\gsim10^{14}~{\rm cm^{-2}}$, where C and/or Si are detected, the
inferred metallicities (in solar units) are $\sim10^{-2.5}$ for C, and
$\sim10^{-2}$ for Si (Songaila \& Cowie 1996, Tytler~et~al.~1995).  There
appears to be no trend of a decreasing mean C/H or Si/H with decreasing
column density, across six orders of magnitude in $N_{\rm HI}$.  Hence, a
plausible assumption is that the metals are universally mixed into the IGM
prior to $z=3$, so that the metallicities of the low column density
absorbers are the same as that found in systems with $N_{\rm
HI}\gsim10^{14}~{\rm cm^{-2}}$.  However, as pointed out by Songaila
(1997), it is also possible that there are no metals in the low column
density systems; in that case, a conservative lower limit for the average
metallicity is $Z_{\rm IGM}=10^{-3}{\rm Z_\odot}$.  The resulting range of
IGM metallicities, $10^{-3}<(Z_{\rm IGM}/{\rm Z_\odot})<10^{-2}$,
translates to the corresponding range $0.017<f_{\rm star}\equiv M_{\rm
star}/M_{\rm gas}<0.17$ for the star formation efficiency.

Figure~\ref{fig:panels} summarizes the resulting reionization
histories of stars or quasars in our $\Lambda$CDM cosmology. The
results for stars are shown in two cases, one with $Z_{\rm
IGM}=10^{-2}{\rm Z_\odot}$ and the other with $Z_{\rm IGM}=10^{-3}{\rm
Z_\odot}$, to bracket the metallicity range mentioned above.  The
assumed IGM metallicity does not affect the quasar calculation.  The
upper left panel of Figure~\ref{fig:panels} shows the fraction of
baryons which reside in collapsed objects with a mass above $M_{\rm
min}=10^{8}{\rm M_{\odot}}[(1+z)/10]^{-3/2}$.  The upper right panel
shows the evolution of the average flux, $J_{21}$, at the local Lyman
limit frequency, in units of $10^{-21}~{\rm
erg~s^{-1}~cm^{-2}~Hz^{-1}~sr^{-1}}$ for quasars (solid line) and
stars (dashed line for $Z_{\rm IGM}=10^{-2}{\rm Z_\odot}$; and dotted
line for $Z_{\rm IGM}=10^{-3}{\rm Z_\odot}$).  The equations used to
obtain $J_{21}$ are the same as described in LH97, except for the
changes in the line element, time--redshift relation, and baryonic
collapsed fraction, caused by our switch to a $\Lambda$CDM
cosmology. The stellar mass function is taken to have the Scalo (1986)
form.  We have assumed that the flux between the local Ly$\alpha$ and
Lyman--limit frequencies vanishes before reionization, due to the
large Gunn--Peterson optical depth.  Although this is a good
approximation for the present purpose, we note that the actual
spectrum in this frequency range is expected to have a sawtooth shape,
as described in Haiman, Rees \& Loeb (1997).  After reionization
$J_{21}$ rises rapidly, and by a redshift $z=3$ the stellar $J_{21}$
reaches a value in the range 0.04--0.4, while quasars contribute
$J_{21}\approx 0.1$.  These values are consistent with the background
flux level inferred from the proximity effect around quasars (Bechtold
1994).

The lower left panel of Figure 7 shows the resulting evolution of the
ionized fraction of hydrogen, $F_{\rm HII}$.  The dashed curves
indicate that for the high end of the allowed metallicity range, stars
ionize the IGM by a redshift $z\sim13$; while the dotted curve shows
that for the lowest metallicity, reionization is delayed until $z=9$.
(Note that for $Z_{\rm IGM}=0.01{\rm Z_\odot}$, HL97 obtained stellar
reionization at a redshift $z=18$ in standard CDM. The change to
$z=13$ here results from our switch to a $\Lambda$CDM cosmology.)  The
solid curve shows that quasars reionize the IGM between these two
redshifts, at $z=11.5$.  This result can be understood simply in terms
of the total number of ionizing photons produced in each case per unit
halo mass: the relative ratios of this number in the three cases are
$1\div0.37\div0.1$, respectively.

Although the reionization redshift can be inferred from the
corresponding damping of CMB anisotropies (see below), it is not
possible to determine whether reionization was caused by stars or
quasars, based on the measurement of this damping alone.  However, the
distinction between the two possibilities could be made if the
reionization redshift of HeII is also measured.  If a source at a
redshift prior to reionization is discovered, the location of the
Gunn--Peterson trough in its spectrum would yield the reionization
redshift, while the shape of the damping wing of the trough can be
used to infer the Gunn--Peterson optical depth (Miralda-Escud\'e
1997). Since the ionization threshold of HeII is four times higher
than that of HI, the ionization of HeII requires a much harder
spectrum.  The relevant measure is $\eta\equiv(dN_{E>54.4{\rm
eV}}/dt)/(dN_{E>13.6{\rm eV}}/dt)$, the ratio of photon fluxes above
54.4eV and 13.6eV.  Equation~($\ref{eq:front}$) reveals that if
recombinations are negligible, and if $\eta\leq n_{\rm He}/n_{\rm
H}\approx 0.08$, then the HeII/HeIII ionization front will lag behind
the HI/HII-ionization front (see Miralda-Escud\'e \& Rees 1994, for a
discussion).  If recombinations are important (redshifts $z\gsim15$),
then the lag is increased further, since HeII recombines 5.5 times
faster than HI.  In our models, the HeII reionization history can be
evaluated by replacing $n_{\rm H}$ with $n_{\rm He}$, and
$dN_{E>13.6{\rm eV}}/dt$ with $dN_{E>54.4{\rm eV}}/dt$, and
multiplying $\alpha_{\rm B}$ by a factor of 5.5, in
equation~(\ref{eq:front}).

We find that for our quasar spectrum, HI and HeII reionization occurs
almost simultaneously (the HeII reionization redshift is $z=11.8$),
but for our stellar spectrum, HeII reionization is never achieved.
These results are explained by fact that for the quasar spectral
template of Elvis~et~al.~(1994), $\eta=0.09$; while for the stellar
spectral template, $\eta\sim10^{-6}$, i.e. there are almost no
HeII--ionizing photons, because most of these photons are absorbed
already in the stellar atmospheres (see Fig.~3 in HL97).  It therefore
seems likely that intergalactic HeII was reionized by quasars rather
than by stars. Thus, the reionization history of HeII provides an
ideal probe for the formation history of the first quasars.

Our results indicate that the reionization redshifts of HI and HeII
should not be very different, unless we have substantially
overestimated the emission from high--redshift quasars, or
underestimated the abundance of high mass stars.
Reimers~et~al.~(1997) have recently claimed a detection of the HeII
reionization epoch at $z=3$.  Their claim is based on the lack of any
detected flux above the redshifted HeII Ly$\alpha$ frequency, although
the flux at the corresponding redshifted HI Ly$\alpha$ frequencies is
transmitted without any absorption.  This observation, however, only
places the limit $\tau_{\rm He}\ga$ few on the HeII Gunn--Peterson
optical depth, rather than the $\tau_{\rm He}\ga$ few thousand,
required to prove HeII reionization.  In fact, Miralda--Escud\'e
(1997) argued that these observational results are likely to have been
caused by spatial variations in the gas density, and fluctuations of
the background radiation intensity above the HeII edge, rather than
the fact that the HeIII bubbles have not yet overlapped.  In summary,
the interpretation of this, and other recent detections of HeII
absorption (Jacobsen~et~al.~1994; Tytler~et~al.~1995;
Davidsen~et~al.~1996; Hogan~et~al~1997) are still controversial, in
that these measurements are consistent with the absorption arising
either from non-overlapping HeIII bubbles, or from other effects
associated with intervening discrete systems (see, e.g. Madau \&
Meiksin 1994; Giroux \& Shull 1997).

\section{Signatures Imprinted on the Cosmic Microwave Background}

Reionization results in a reduction of the temperature anisotropies of
the cosmic microwave background (CMB).  The free electrons released
during this epoch increase the optical depth of the universe to
electron scattering ($\tau_{\rm es}$), and damp the CMB anisotropies
on scales below the size of the horizon at that time, $\la 10^\circ$
(Efstathiou \& Bond 1987; Kamionkowski, Spergel, \& Sugiyama 1994).
The electron scattering optical depth in our models is approximately
$\tau_{\rm es}(z)\approx0.053\Omega_{\rm b} h\int dz \sqrt{1+z} F_{\rm
HII}(z)$, where $F_{\rm HII}(z)$ is the filling factor of the HII
regions.  Hu \& White (1997) provide a fitting formula for the damping
factor of the CMB power--spectrum, $R_\ell^2$, as a function of the
index $\ell$ in the spherical harmonic decomposition of the microwave
sky [the angular scale corresponding to a given $\ell$--mode is
$\sim1^\circ \times (\ell/200)^{-1}$]. The damping factor is uniquely
related to the evolution of the optical depth for electron scattering
as a function of redshift.

The lower right panel of Figure~\ref{fig:panels} shows both $\tau_{\rm
es}(z)$ and $R_{\ell}^2$ for the stellar and quasar reionization histories.
The electron scattering optical depth is 3--5\% for the stellar case, and
$\sim$4\% for the quasar case.  The corresponding damping factors for the
power spectrum of CMB anisotropies are 6--10\%, and $\sim$8\%,
respectively.  Although the amplitude of these damping factors is small,
they are within the proposed sensitivities of the future MAP and Planck
satellite experiments, if data on both temperature and polarization
anisotropies of the CMB will be gathered (see Table 2 in Zaldarriaga et al.
1997).

The UV emission from early quasars could also distort the spectrum of the
CMB, after being reprocessed by dust.  An early epoch of star formation and
metal enrichment is inevitably accompanied by the formation of dust in
supernova shells.  This dust would absorb the UV flux from both stars and
quasars, and re--emit it at longer wavelengths.  The re--emitted radiation
is added to the CMB spectrum, and introduces a deviation from its perfect
blackbody shape.  A distortion of this type due to population III stars
alone has been considered by several authors (Wright et al. 1994; Bond,
Carr \& Hogan 1991; Adams et al. 1989; Wright 1981).

In a previous paper (LH97), we have calculated the distortion
amplitude from early stars in a standard CDM cosmology, using the
Press-Schechter formalism and normalizing the star formation
efficiency in collapsed objects based on the observed metallicity of
the IGM.  The essential assumptions in this calculation were: (1) the
initial mass function (IMF) of stars at high--redshifts is the same as
observed locally (Scalo 1986), and (2) each type II supernova yields
${\rm 0.3M_\odot}$ of dust, which gets uniformly distributed within
the intergalactic medium.  We further assumed that the dust is in
thermal equilibrium with the total (CMB + stellar) radiation field,
that its absorption follows the wavelength-dependent opacity of
Galactic dust (Mathis 1990), and that it radiates as a blackbody at
its equilibrium temperature.  We have found that allowing two dust
components made of graphite and silicate (Draine \& Lee 1984) with
independent equilibrium temperatures, does not change our results by
more than a few percent.  For a range of possible parameter values
within standard CDM, we found that the opacity of intergalactic dust
to infrared sources at redshifts $z\ga 10$ is significant, $\tau_{\rm
dust}= (0.1$--$1)$, and that this dust distorts the microwave
background spectrum by a Compton $y$--parameter in the range
$(0.06$--$6)\times 10^{-5}$.

Here, we repeat the LH97 calculations in a $\Lambda$CDM cosmology, and
extend them by adding the UV flux from quasars to the total radiation
field.  Our new results are presented in Figure~\ref{fig:dust}.  The
top panel of this figure shows the resulting total spectrum of the
radiation background (CMB + direct stellar and quasar emission + dust
emission) at $z=3$. The spectrum describes a minimum level of CMB
distortion down to $z=3$; more distortion could be added between
$0<z<3$ by dust and radiation from galaxies.  However, the additional
emission at $0<z<3$ does not relate directly to the reionization epoch
and is therefore ignored here.  The dashed lines show the spectrum for
the stellar case, for $Z_{\rm IGM}=10^{-2}{\rm Z_\odot}$ (top curve)
and $Z_{\rm IGM}=10^{-3}{\rm Z_\odot}$ (bottom curve).  The pronounced
peak at a wavelength of $\sim1\mu$m is the sum of the direct starlight
from high--z stars, redshifted to $z=3$.  The deviation from the pure
$T_{\rm CMB}=2.728(1+z)$K blackbody shape (shown by the dotted curve)
are visible in this figure below a wavelength of 0.02cm. The
corresponding distortions in the COBE range is quantified by the
Compton $y$--parameter, defined as \beq y_{\rm c}\equiv
\frac{1}{4}\left[\frac{\int d\nu N_{\nu}}{\int d\nu N_{0,\nu}}-1
\right], \eeq where the integrals are evaluated over the FIRAS
frequency range of 60--600 GHz (Fixsen et al. 1996), $N_{\nu}$ is the
comoving number density of photons with a comoving frequency $\nu$ in
the total radiation field, and $N_{0,\nu}$ reflects the unperturbed
CMB,
\beq 
N_{0,\nu}\equiv\frac{8\pi}{c^3}
\frac{\nu^3}{\exp(h\nu/k_{\rm B}T_{\rm CMB})-1}. \label{eq:10} 
\eeq
The bottom panel of Figure~\ref{fig:dust} shows the redshift evolution of
the $y$--parameter, as well as the dust temperature, $T_{\rm dust}$.  For
the stellar cases considered above, we obtain $1.3\times10^{-7}<y_{\rm
c}<1.2\times10^{-5}$ at $z=3$ (dashed lines).  Since both the dust opacity
and the stellar radiation background are independently proportional to the
star--formation efficiency $f_{\rm star}$, the spectral distortion is
proportional to $f_{\rm star}^2$ (see LH97).  For the high end of the
allowed metallicity range, the distortion at $z=3$ is just below the upper
limit set by COBE, $y=1.5\times10^{-5}$ (Fixsen~et~al.~1996); for lower
metallicities, the $y$--parameter drops rapidly below this value.  The dust
temperature remains in all cases close the the CMB temperature (shown by
the dotted line).

The solid lines in Figure~\ref{fig:dust} show how these results change
when the radiation from the early quasars is added. In particular, the
top panel shows that the broad emission spectrum of quasars
(cf. Fig. 1) adds a background radiation field over a much broader
range of wavelengths than stellar emission.  But since dust is
effective in absorbing the quasar flux only in a relatively narrow
range around $\sim 1\mu$m, the distortion in the COBE regime is not
significantly altered.  In particular, we find
$4.1\times10^{-6}<y_{\rm c}<2.0\times10^{-5}$ at $z=3$ for the range
$10^{-3}{\rm Z_\odot}<Z_{\rm IGM}<10^{-3}{\rm Z_\odot}$.  It is
interesting to note that the long-wavelength emission from high
redshift quasars in the COBE wavelength range, results in a minimum
distortion of $y_{\rm c}\approx 3.4\times10^{-6}$.  Depending on
$Z_{\rm IGM}$, this amounts to a fraction $\sim17$--$82$\% of the
total $y$--parameter values quoted before.  This minimal level of
distortion results simply from the cumulative far-infrared flux of
early quasars, and should be present even in the absence of any
intergalactic dust.  It can be avoided only if early quasars have much
less emission longward of $\sim 100\mu$m, relative to the spectrum
shown in Figure 1.

\section{Number Counts of Faint Quasars}

Next we examine the feasibility of direct detection of the early
population of low-luminosity quasars.  Although any such detection is
beyond the sensitivities of current observing programs, future
instruments such as the Space Infrared Telescope Facility (SIRTF) or
the Next Generation Space Telescope (NGST), will achieve the required
sensitivity.  NGST, which is scheduled for launch a decade from now,
is expected to reach a sensitivity of $\sim 1$ nJy in the wavelength
range 1--3.5$\mu$m (Mather \& Stockman 1996; see also
http://ngst.gsfc.nasa.gov).  The number of sources that NGST would
detect at redshifts $z>z_{\rm min}$, per unit solid angle, with
observed flux between $F_\nu$ and $F_\nu+dF_\nu$ (averaged over the
observed wavelength range 1--3.5$\mu$m), is given by
\beq 
\frac{dN}{d\Omega dF_\nu}(F_\nu,z_{\rm min})=
\int_{z_{\rm min}}^{\infty} dz 
\left(\frac{dV_{\rm c}}{dzd\Omega}\right) n_{\rm c}(z,F_\nu),
\eeq
where $dV_{\rm c}/dzd\Omega$ is the comoving volume element per unit
redshift per unit solid angle, and $n_{\rm c}(z,F_\nu)$ is the
comoving number density of objects at redshift $z$, whose flux at
$z=0$ is observed to be between $F_\nu$ and $F_\nu+dF_\nu$.  Based on
the halo formation rate $d^2N_{\rm ps}/dMdz$
[cf. Eq.~(\ref{eq:rate})], $n_{\rm c}$ is given by a sum over halos of
different ages that exist at each redshift,
\beq
n_{\rm c}(z,F_\nu)=\int_z^{\infty} dz^\prime
\frac{dM}{dF_\nu}(z,z^{\prime},F_\nu)
\left.\frac{d^2N_{\rm ps}}{dMdz^\prime}
\right|_{M(z,z^{\prime},F_\nu),z^\prime}, \label{eq:ncom}
\eeq
where the factor $dM/dF_\nu$ converts the number density per unit mass
interval to number density per unit flux interval, based on
equation~(\ref{eq:haloflux}).  The assumptions implicit in this
expression are that the simple derivative $d^2N_{\rm ps}/dMdz$ gives
the halo formation rate [see discussion following
eq.~(\ref{eq:rate})], and that the halos do not grow in mass during
the active phase of the emission sources they host (a good assumption
for the short--lived quasar activity).  Relaxing any of these
assumptions would increase our predicted number counts.  Note that, in
HL97 we had effectively identified $z=z^{\prime}$ in the kernel of
equation~(\ref{eq:ncom}), when estimating the number counts of star
clusters. This might be justified only for stellar sources that live
for a Hubble time or longer.  However, in this case there is a danger
of double counting merging halos in neighboring redshift bins
(although at sufficiently high redshifts objects are exponentially
rare and their merging probability is small). We intend to address
this issue more rigorously using the excursion-set formalism (Bond et
al. 1991; Lacey \& Cole 1993, 1994), in a future publication.

The average flux $F_\nu(z,M)$ from a halo at
redshift $z$ with mass $M$, that was formed at $z^\prime\ge z$ equals
\beq
F_\nu(z,z^\prime)=\frac{10^{32}}{d_{\rm L}(z)^2}
\frac{\epsilon M}{\Delta\nu_0} \int_{\nu_{\rm min}}^{\nu_{\rm max}}
d\nu j(\nu,t_{z,z^\prime})~~~~~~{\rm nJy}.
\label{eq:haloflux}
\eeq
Here the halo mass $M$ is in units of ${\rm M_\odot}$; $\nu_{\rm min}$
and $\nu_{\rm max}$ are the redshifted frequencies (in units of Hz)
corresponding to the observed wavelengths of 3.5$\mu$m and 1$\mu$m,
respectively; $\Delta\nu_0=2.14\times10^{14}$ Hz is the width of the
NGST frequency band; $d_{\rm L}(z)$ is the luminosity distance in cm;
$\epsilon=10^{-3.2}$ for quasars or $\epsilon=0.17\Omega_{\rm
b}/\Omega_{\rm m}=0.02$ for stars, is the fraction of halo mass in the
form of the central black hole or stars, respectively;
$j(\nu,t_{z,z^\prime})$ is the template luminosity per unit (stellar
or black hole) mass for either stars or quasars, in units of ${\rm
erg~sec^{-1}~Hz^{-1}~sr^{-1}~M_\odot^{-1}}$; and $t_{z,z^\prime}$ is
the time elapsed between redshifts $z^\prime$ and $z$. The time
dependence of $j(\nu)$ is obtained from our fitted light--curve
function $f(t)$ in equation~(\ref{eq:lightcurve}) for quasars, and
from tabulated evolutionary tracks for stars (Schaller~et~al.~1992;
see HL97 for details).  We suppressed the emergent flux from all
objects at emission frequencies above \lya~before reionization occurs
(assuming sudden reionization at the redshift $z=13$ or $9$ for stars,
and $z=11.5$ for quasars), since the Gunn--Peterson optical depth of
HI at these frequencies is exceedingly high prior to reionization.
 
Figure~\ref{fig:ncounts} shows the predicted number counts normalized
to the field of view of SIRTF (which is scheduled for launch long
before NGST), $5^{\prime}\times5^{\prime}$; the planned field-of-view
of NGST is similar, $4^\prime\times4^\prime$.
Figure~\ref{fig:ncounts} shows separately the number per logarithmic
flux interval of all objects with $z>5$ (thin lines), and with $z>10$
(thick lines).  In general, the number of detectable sources is high.
With its expected sensitivity, NGST will be able to probe about
$10^{3}$ quasars at $z>10$, and $10^{3.5}$ quasars at $z>5$ (solid
curves) per field of view.  The bright--end tail of the number counts
approximately follows the power law $dN/dF_\nu\propto F_\nu^{-2.5}$.
This slope is too shallow for foreground galaxies to significantly
increase the number counts due to the amplification bias from
gravitational lensing.  The top solid curve also shows that the number
of faint quasars starts to turn over below $\sim1$ nJy, as a result of
the cutoff in the minimum halo mass that allows atomic line cooling.
The average angular separation between the quasars at $z>10$ would be
$\sim$10\H{}, well above the angular resolution limit of 0.06\H{}
planned for NGST.  For comparison, we show in Figure~\ref{fig:ncounts}
the corresponding number counts of ``star--clusters'', i.e. assuming
that each halo shines due to a starburst that converts a fraction
0.017-0.17 of the gas into stars.  The dashed lines indicate that NGST
might detect $10^{1.5}$--$10^{3.5}$ star--clusters at $z>10$ per field
of view, and $10^{3.5}$--$10^{4.5}$ clusters at $z>5$.  Unlike
quasars, star clusters could in principle be resolved, if they extend
over a scale comparable to the virial radius of the collapsed halo
(Haiman \& Loeb 1997b).

\section{Conclusions}

\begin{figure}[b]
\vspace{2.6cm}
\includegraphics{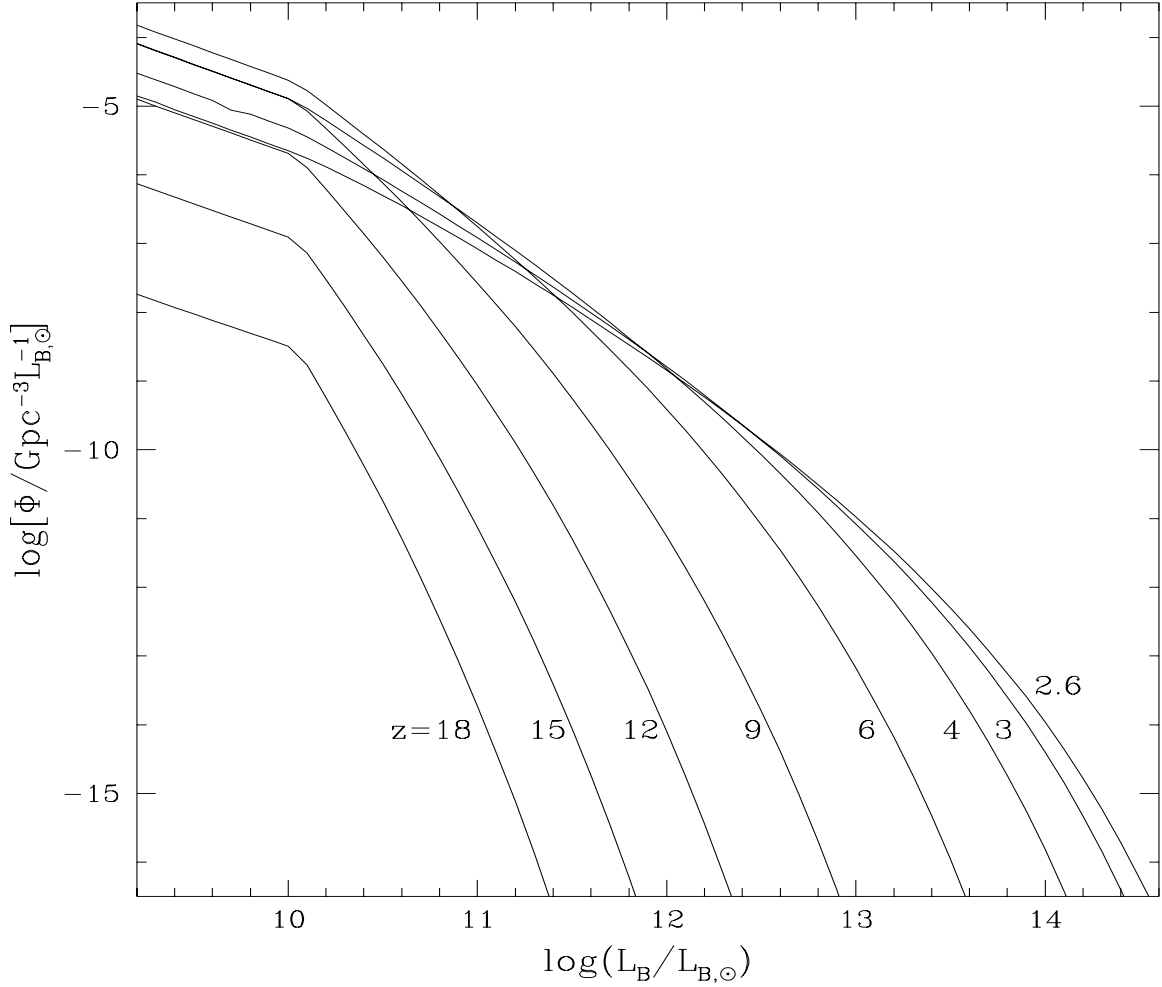}
\vspace*{4.5in}
\caption[LFs ] {\label{fig:lfs} The extrapolated quasar luminosity
function at high redshifts, using the light--curve of
Figure~\ref{fig:lcurve}.  The break at $L_{\rm B}\sim10^{10}{\rm
L_{B,\odot}}$ is a result of the low mass cutoff for the dark matter halo,
$M_{\rm min}=10^{8}{\rm M_{\odot}}[(1+z)/10]^{-3/2}$, that defines the
threshold for efficient (atomic) cooling of virialized gas.}
\end{figure}

We extrapolated the quasar luminosity function to faint magnitudes and
high redshifts, based on the assumptions that the halo formation rate
follows the Press--Schechter theory, and that the quasar light--curve
scales linearly with the halo mass, and is otherwise a universal
function of time.  We demonstrated that a universal light--curve of
the form, $L(t)=L_{\rm Edd}\exp(-t/t_0)$, with $t_0=6.6\times 10^5$ yr
and $L_{\rm Edd}$ being the Eddington luminosity, provides an
excellent fit to observational data for the quasar LF between
$2.6<z<4.5$ provided that the final black hole mass is a fraction
$\sim 5.4\times 10^{-3}$ of the baryonic mass in each object.  This
fraction is surprisingly close to the roughly universal ratio between
black hole and bulge masses observed in nearby galaxies (Magorrian et
al.  1997).  Our extrapolation of the quasar LF to high redshifts and
low luminosities as presented in Figure~\ref{fig:lfs}, can be viewed
as a natural by--product of this physically motivated fit.

\begin{figure}[b]
\vspace{2.6cm}
\includegraphics{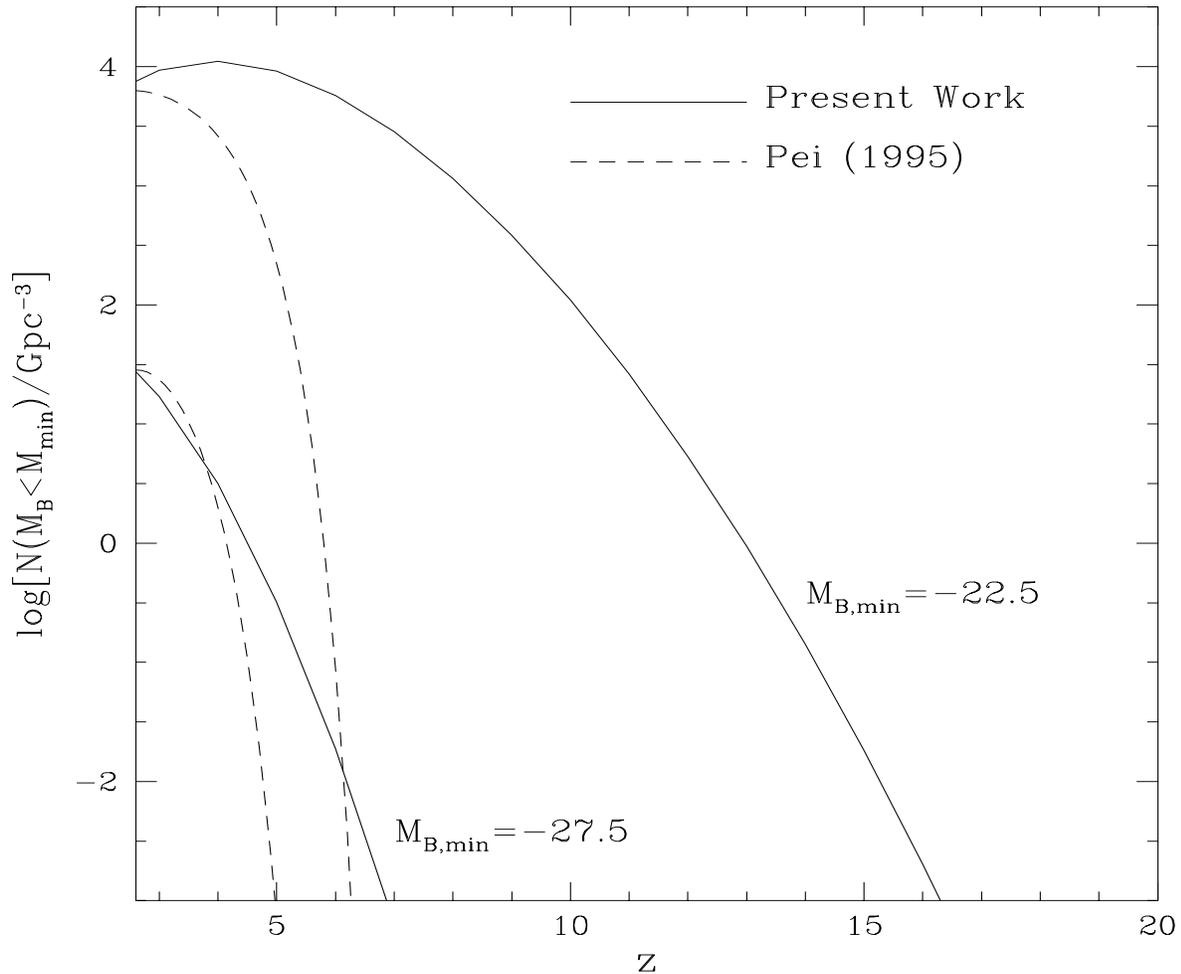}
\vspace*{4.5in}
\caption[Number Density of Quasars] {\label{fig:nz} The redshift
evolution of the number density of quasars.  Solid lines show the
predictions from our model, while the dashed lines show the formal
extrapolation of the fitting formulae obtained by Pei (1995).}
\end{figure}

\begin{figure}[b]
\vspace{2.6cm}
\includegraphics{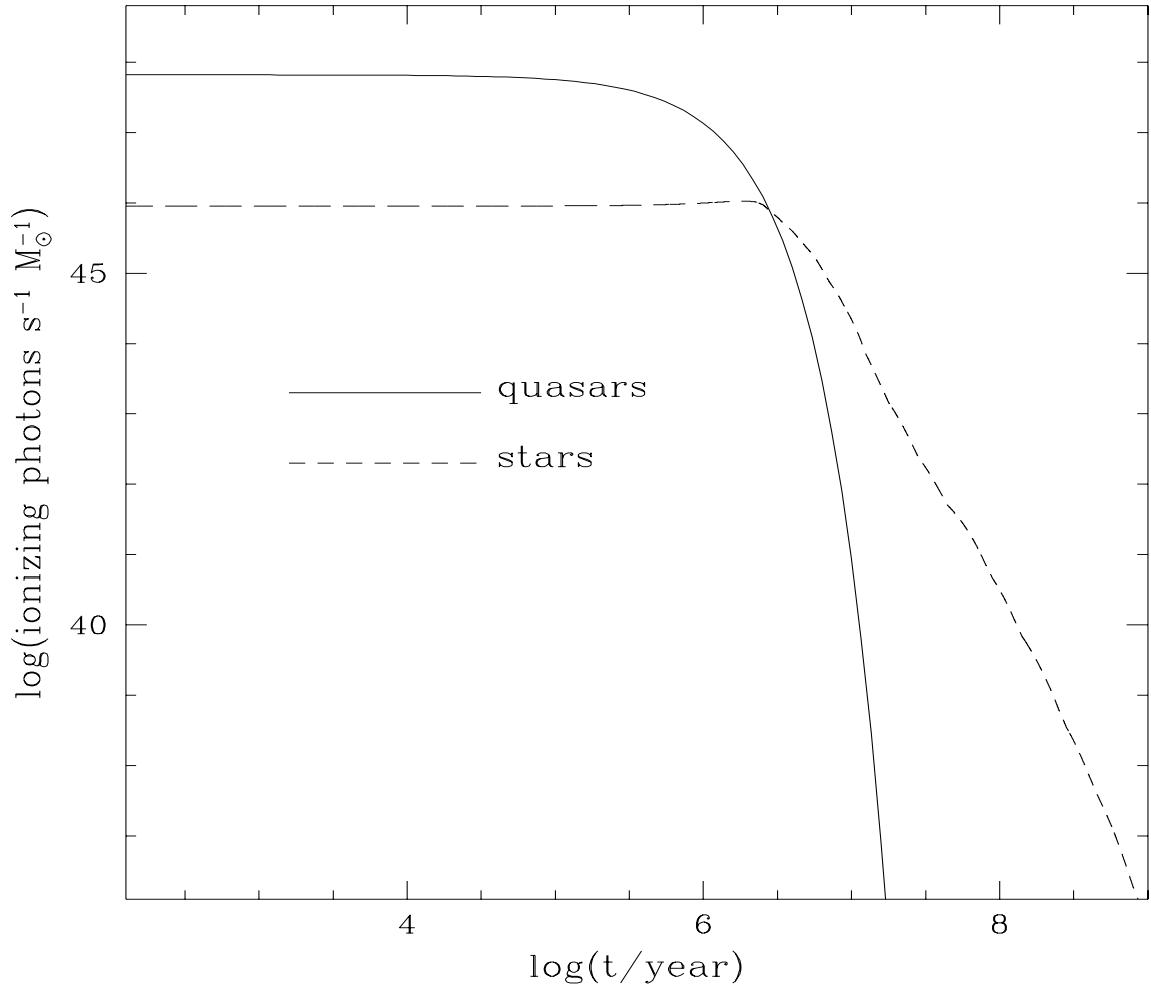}
\vspace*{4.5in}
\caption[Ionizing Flux] {\label{fig:nion} Evolution of the production
rate of ionizing photons, per unit black hole mass in quasars (solid
curve), and per unit stellar mass in a starburst (dashed curve).}
\end{figure}

\begin{figure}[b]
\vspace{2.6cm}
\includegraphics{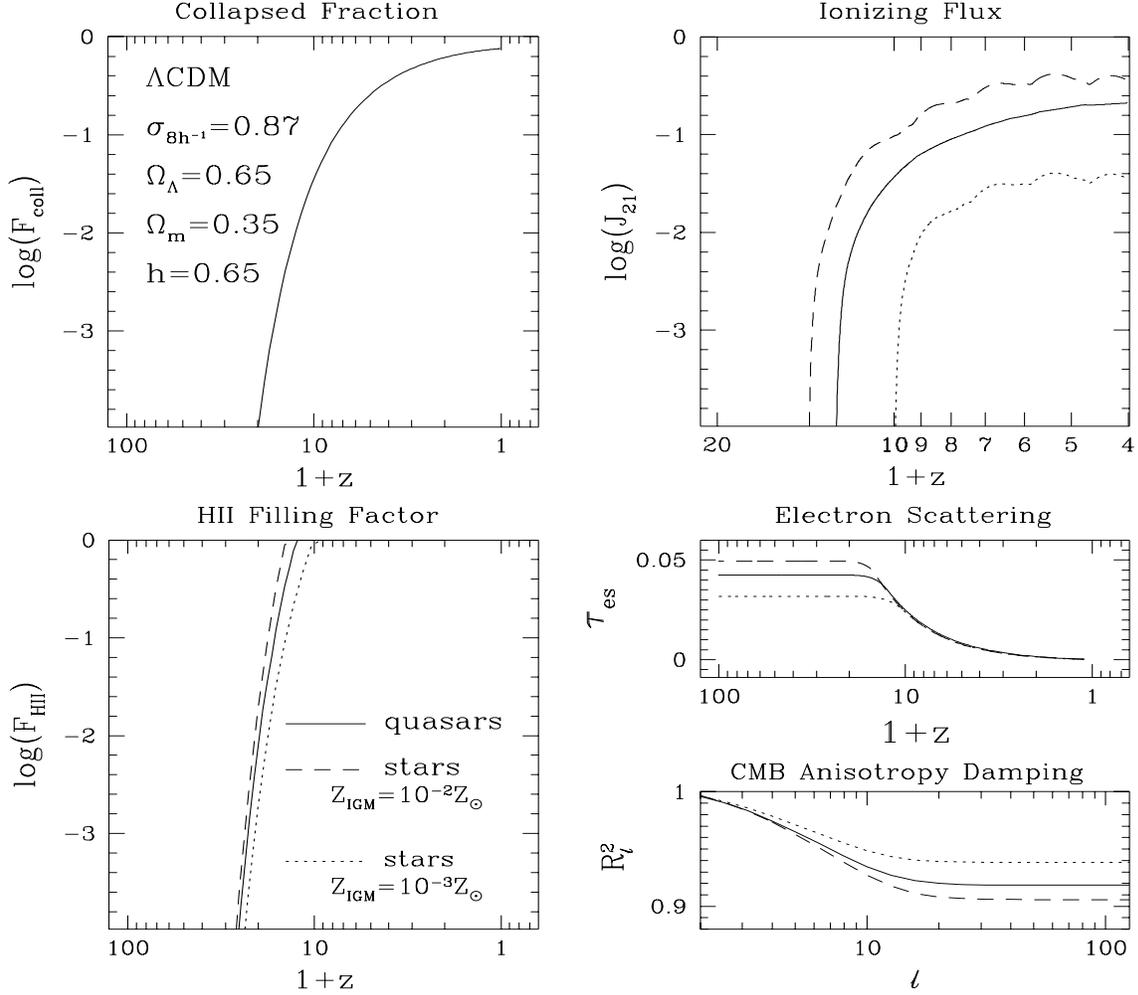}
\vspace*{4.5in}
\caption[Reionization] {\label{fig:panels} Reionization history from
quasars (solid lines) and early stars (dashed [${\rm Z_{IGM}=
10^{-2}Z_\odot}$] and dotted [${\rm Z_{IGM}=10^{-3}Z_\odot}$] lines).
The different panels show (i) the collapsed fraction of baryons; (ii)
the evolution of the flux at the local Lyman limit frequency, in units
of $10^{-21}~{\rm erg~s^{-1}~cm^{-2}~Hz^{-1}~sr^{-1}}$; (iii) the
volume filling factor of ionized regions; and (iv) the optical depth
to electron scattering, $\tau_{\rm es}$, and the corresponding damping
factor, ${\rm R^2_{\ell}}$, for the power--spectrum decomposition of CMB
anisotropies as a function of the spherical harmonic index $\ell$.}
\end{figure}

\begin{figure}[b]
\vspace{2.6cm}
\includegraphics{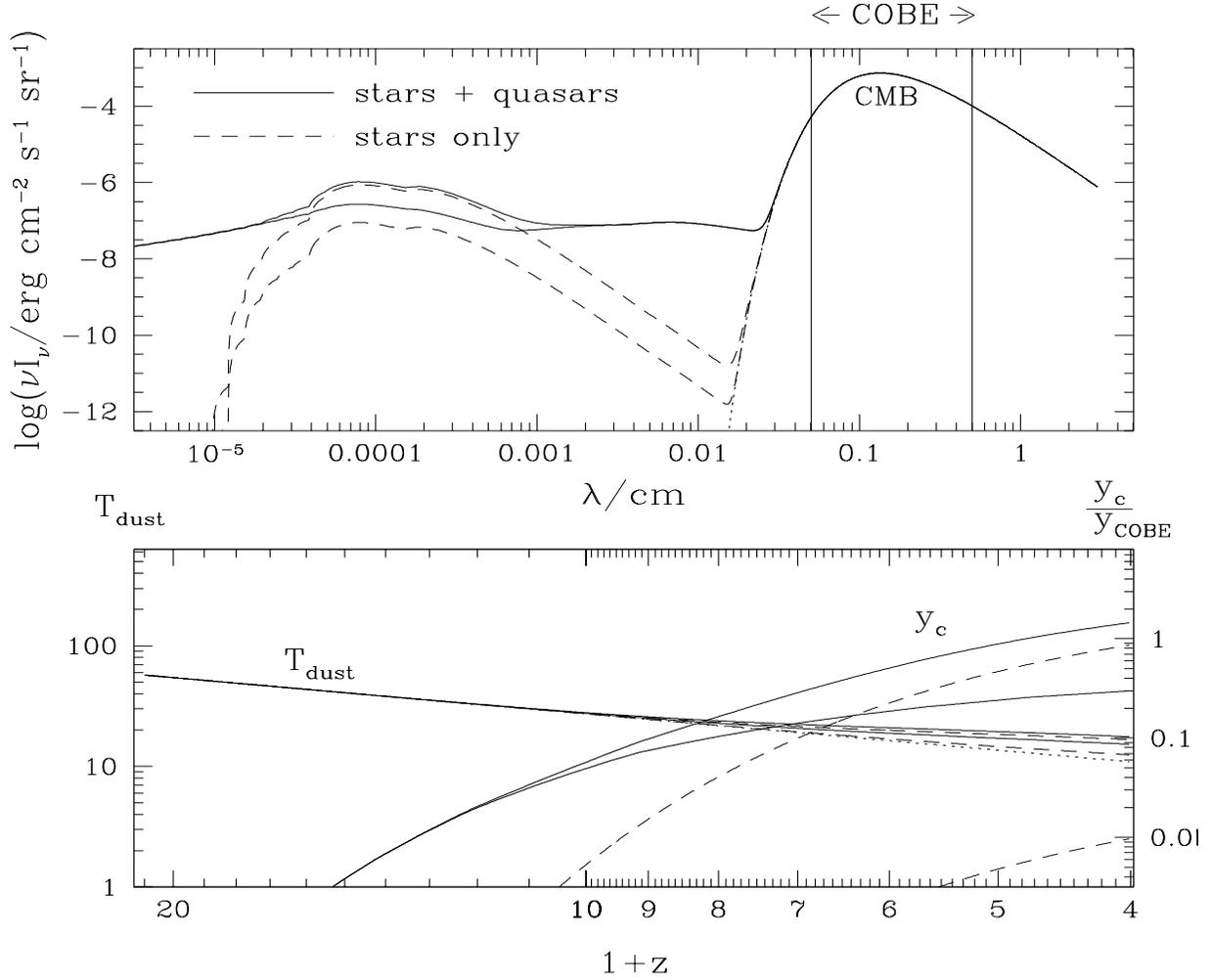}
\vspace*{4.5in}
\caption[Dust] {\label{fig:dust} Upper panel: The full (CMB + direct
light + dust emission) background flux at redshift $z=3$ in the
presence of early stars alone (dashed lines) and in the presence of
both stars and quasars (solid lines).  The pairs of dashed and solid
lines bracket the possible range of the IGM metallicity, $10^{-3}{\rm
Z_\odot}<Z_{\rm IGM}<10^{-2}{\rm Z_\odot}$.  Lower panel: The
corresponding redshift evolution of the dust temperature, and the
Compton $y$--parameter.  The dotted curve shows the CMB temperature,
$T_{\rm CMB}=2.728(1+z)$K.}
\end{figure}

\begin{figure}[b]
\vspace{2.6cm}
\includegraphics{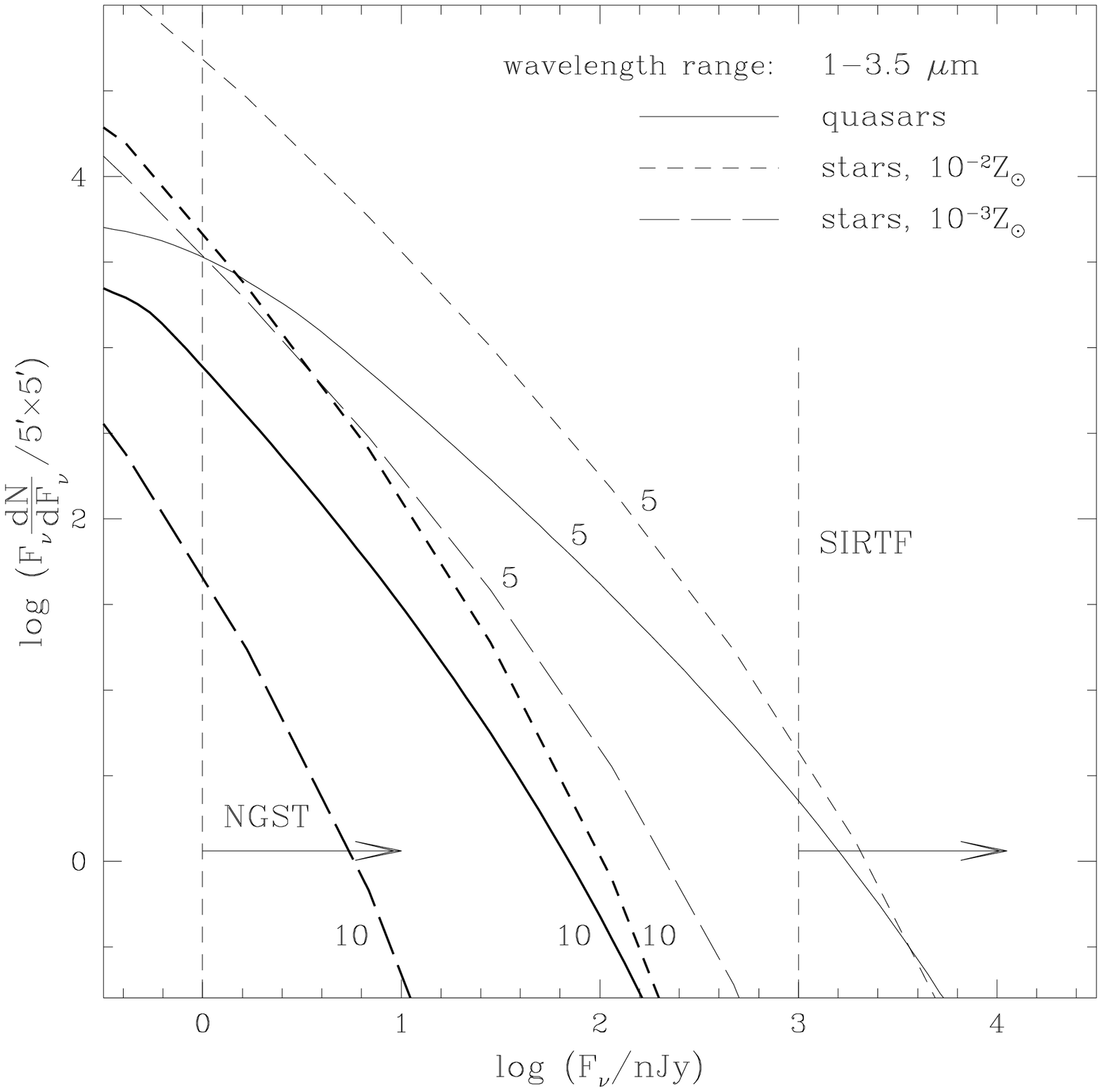}
\vspace*{4.5in}
\caption[Number Counts] {\label{fig:ncounts} The number per
logarithmic flux interval of high--redshift objects which could be
probed by future space telescopes, in the wavelength range of
1--3.5$\mu$m.  We assume sudden reionization at $z$=11.5 for quasars,
and at $z$=9 or 13 for stars.  The vertical dashed lines show the
expected sensitivities of the Space Infrared Telescope Facility
(SIRTF) and the Next Generation Space Telescope (NGST).  The thick
lines, labeled ``10'', correspond to objects located at redshifts
$z>10$, and the thin lines, labeled ``5'', correspond to objects located
at redshifts $z>5$.}
\end{figure}

Given the Press-Schechter prediction for the formation rate of halos,
and the fact that $t_0$ is much shorter than the Hubble time, our
prediction for the reionization history depends only on the integral
of the inferred light--curve, namely on the mean radiative efficiency
of the material that made the black hole.  The low value inferred for
$t_0$ implies an average radiative efficiency of only $\sim 0.1$\%.
This rather low estimate for the efficiency is smaller by two orders
of magnitude than the value expected in thin disk accretion; any
increase in its value would only enhance the effects we
predict. Despite our conservative approach, we predict an extensive
population of faint quasars at high redshifts.  For illustration, we
show in Figure~\ref{fig:nz} the expected evolution in the number
density of bright quasars with absolute $B$ magnitude $M_B<-27.5$,
along with that of quasars which are two orders of magnitude fainter.
Although the density of bright quasars in our models declines rapidly
beyond redshift $z\sim 3$ in good agreement with existing observations
at $2.6<z<4.5$, we predict an abundant population of faint quasars
which extends out to redshifts $z\ga 10$, and has not yet been
directly observed.

We find that, in general, the effects of early quasars are comparable
to those of early stars.  However, stellar effects scale with the
average IGM metallicity that we used to normalize the star formation
efficiency.  If the observed values of the C/H ratio in the individual
Ly$\alpha$ forest clouds are representative of the universal average
C/H ratio, than this ratio implies an IGM metallicity of $\sim
0.5$--1\% solar.  However, if the overall carbon abundance is patchy,
because it is not well mixed with the rest of the intergalactic gas
(as argued by Gnedin 1997), then the true average metallicity may be
an order of magnitude lower, 0.1\% solar (Songaila 1997).  We have
found that at the high end of the allowed metallicity range, the
effects of stars are stronger than that of quasars, while at the low
end of this range, quasars are more effective.  In particular, quasars
reionize the universe at $z\sim11.5$, while stars at $9\lsim
z\lsim13$.  The Compton $y$--parameter for the spectral distortion of
the CMB is expected to be $1.3\times10^{-7}<y_{\rm
c}<1.2\times10^{-5}$ from starlight alone, and
$4.1\times10^{-6}<y_{\rm c}<2.0\times10^{-5}$ if the flux of quasars
is added. The direct far-infrared flux from quasars yields a lower
limit to the distortion, $y_{\rm c}\ga 3.4 \times 10^{-6}$, even in
the absence of any intergalactic dust. The expected number of $z>10$
quasars per square arcminute, brighter than 1nJy in the wavelength
range 1--3.5$\mu$m, is $\sim40$, while the corresponding number for
star clusters is $\sim1$--$100$. Both of these numbers are large
enough to show as a distinct population on a typical image by the
forthcoming Next Generation Space Telescope.

\acknowledgements

We thank J. McDowell, Y. Pei, and W. Zheng for data on quasar spectra
and the quasar luminosity function, and L. Hui, J. Miralda-Escud\'e,
R. Narayan, M. Rees, E. Turner, and A. Songaila for useful
discussions. This work was supported in part by the NASA ATP grant
NAG5-3085, and the Harvard Milton fund.

%############################################################################
%########################### FIGURES ########################################
%############################################################################

%########################## figure 1 ########################################
\clearpage
\newpage

%########################## figure 2 ########################################
\clearpage
\newpage

%########################## figure 3 ########################################
\clearpage
\newpage

%########################## figure 4 ########################################
\clearpage
\newpage

%########################## figure 5 ########################################
\clearpage
\newpage

%########################## figure 6 ########################################
\clearpage
\newpage

%########################## figure 7 ########################################
\clearpage
\newpage

%########################## figure 8 ########################################
\clearpage
\newpage

%########################## figure 9 ########################################
\clearpage
\newpage


\begin{references}

\reference{}
Adams, F. C., Freese, K., Levin, J., McDowell, J. 1989, ApJ, 344, 24
\reference{}
Bechtold, J. 1994, ApJS, 91, 1
\reference{}
Bennett, C. L., Banday, A. J., Gorski, K. M., Hinshaw, G., Jackson, P., 
Keegstra, P., Kogut, A., Smoot, G. F., Wilkinson, D. T., \& Wright, E. L.
1996, ApJ, 464, L1
\reference{}
Bond, J. R. 1995, in Theory and Observations of the Cosmic
Microwave Background Radiation, ed. Schaeffer, R. (Elsevier:
Netherlands), in press
\reference{}
Bond, J. R., Cole, S., Efstathiou, G. \& Kaiser, N. 1991, ApJ, 379, 440
\reference{}
Bond, J. R., Carr, B. J., \& Hogan, C. J. 1991, ApJ, 367, 420
\reference{}
Boyle, B. J., Jones, L. R., Shanks, T., Marano, B., Zitelli, V., \& Zamorani, G. 1991, in Crampton, D., ed., ASP Conf. Series No. 21, The Space Distribution
of Quasars, Astron. Soc. Pacif., San Francisco, p. 191
\reference{}
Carlberg, R. G. 1990, ApJ, 350, 505
\reference{}
Carr, B. J., Bond, J. R., \& Arnett, W. D. 1984, ApJ, 277, 445
\reference{}
Carroll, S. M., Press, W. H., Turner, E. L. 1992, ARA\&A, 30, 499
\reference{}
Davidsen, A. F., Kriss, G. A., Zheng, W. 1996, Nature, 380, 47
\reference{}
Draine, B. T., \& Lee, H. M. 1984, ApJ, 285, 89
\reference{}
Efstathiou, G., \& Bond, J. R. 1987, MNRAS, 227, 33p
\reference{}
Efstathiou, G., \& Rees, M. J. 1988, MNRAS, 230, 5p
\reference{}
Eisenstein, D. J. 1997, ApJ, submitted, preprint astro-ph/9709054
\reference{}
Eisenstein, D.J., \& Loeb, A. 1995, ApJ, 443, 11
\reference{}
Elvis, M., Wilkes, B. J., McDowell, J. C., Green, R. F., Bechtold, J., 
   Willner, S. P., Oey, M. S., Polomski, E., \& Cutri, R. 1994, ApJS, 95, 1
\reference{}
Fixsen, D. J., Cheng, E. S., Gales, J. M., Mather, J. C., Shafer, R. A., 
Wright, E. L. 1996, ApJ, 473, 576
\reference{}
Frank, J., King., A., \& Raine, D. 1992, Accretion Power in Astrophysics
(Cambridge: Cambridge Univ. Press)
\reference{}
Franx, M, Illingworth, G. D., Kelson, D. D., van Dokkum, P. G., \&
Tran, K.-V. 1997, ApJ, 486, L75
\reference{}
Giroux, M. L., \& Shull, J. M. 1997, AJ, 113, 1505
\reference{}
Gnedin, N. Y., MNRAS, in press, preprint astro-ph/9709224
\reference{}
Gnedin, N. Y., \& Ostriker, J. P. 1997, ApJ, 486, 581
\reference{}
Haiman, Z., \& Loeb, A. 1997a, ApJ, 483, 21 (HL97)
\reference{}
---------------------------- 1997b, to appear in the Proceedings of
Science with the Next Generation Space Telescope (Eds. E. Smith \& A. Koreans),astro-ph/9705144
\reference{}
Haiman, Z., Rees, M. J., \& Loeb, A. 1996, ApJ, 467, 522
\reference{}
---------------------------- 1997, ApJ, 476, 458
\reference{}
Haiman, Z., Thoul, A., \& Loeb, A. 1996, ApJ, 464, 523
\reference{}
Haehnelt, M. G., \& Rees, M. J. 1993, MNRAS, 263, 168
\reference{}
Hogan, C. J., Anderson, S. F., Rugers, M. H. 1997, AJ, 113, 1495
\reference{}
Hu, W., \& White, M. 1997, ApJ, 479, 568
\reference{}
Jacobsen, P., Boksenberg, A., Deharveng, J. M., Greenfield, P., Jedrzejewski, R., \& Paresce, F. 1994, Nature, 370, 35
\reference{}
Kamionkowski, M., Spergel, D. N., \& Sugiyama, N. 1994, ApJL, 426, L57
\reference{}
Kormendy, J., Bender, R., Magorrian, J., Tremaine, S., Gebhardt, K.,
   Richstone, D., Dressler, A., Faber, S. M., Grillmair, C., \& Lauer,
   T. R. 1997, ApJ, 482, L139
\reference{}
Lacey, C. G., \& Cole, S. 1993, MNRAS, 262, 627
\reference{}
---------------------------- 1994, MNRAS, 271, 676
\reference{}
Loeb, A. 1997, ``The First Stars and Quasars in the Universe'', to appear
in Proc. of ``Science with the Next Generation Space Telescope'', April
1997, preprint astro-ph/9704290
\reference{}
Loeb, A., \& Haiman, Z. 1997, ApJ, 490, in press, preprint 
astro-ph/9704133 (LH97)
\reference{}
Magorrian, J., et al. 1997, submitted to ApJ, preprint astro-ph/9708072
\reference{}
Madau, P., \& Meiksin, A. 1994, ApJ, 433, L53
\reference{}
Mather, J, \& Stockman, P. 1996, STSci Newsletter, v. 13, no. 2, p. 15
\reference{}
Mathis, J. S. 1990, ARA\&A, 28, 37
\reference{}
McLeod, K. 1996, in ESO-IAC conference on Quasar Hosts,
Tenerife, in press
\reference{}
Miralda-Escud\'e, J. 1997, ApJ, in press, preprint astro-ph/9708253
\reference{}
Miralda-Escud\'e, J., \& Rees, M. J. 1994, MNRAS, 266, 343
\reference{}
---------------------------- 1997, ApJ, 478, L57
\reference{}
Narayan, R. 1996, ``Advective Disks'', to appear in Proc. IAU Colloq. 163
on Accretion Phenomena \& Related Outflows, A.S.P. Conf. Series, eds. D. T.
Wickramasinghe, L.  Ferrario, G. V. Bicknell, in press, preprint
astro-ph/9611113
\reference{}
Ostriker, J. P., \& Steinhardt, P. J. 1995, Nature, 377, 600
\reference{}
Pei, Y. C. 1995, ApJ, 438, 623
\reference{}
Press, W. H., \& Schechter, P. L. 1974, ApJ, 181, 425
\reference{}
Rees, M. J. 1996, preprint astro-ph/9608196
\reference{}
Reimers, D., K\"ohler, S., Wisotzki, L., Groote, D., Rodriguez-Pascual, P., 
   \& Wamsteker, W. 1997, A\&A, in press, preprint astro-ph/9707173
\reference{}
Sasaki, S. 1994, PASJ, 46, 427
\reference{}
Scalo, J. M. 1986, Fundamentals of Cosmic Physics, vol. 11, p. 1-278
\reference{}
Schaller, G., Schaerer, D., Meynet, G., \& Maeder, A. 1992, A\&ASS, 96, 269
\reference{}
Schneider, D. P., Schmidt, M., \& Gunn 1991, J. E, AJ, 102, 837
\reference{}
Scott, D., Silk, J., \& White, M. 1995, Science, 268, 829
\reference{}
Shapiro, P. R., \& Giroux, M. L. 1987, ApJ, 321, L107
\reference{}
Small, T. A., \& Blandford, R. D. 1992, MNRAS, 259, 725
\reference{}
Songaila, A. 1997, ApJL, in press, preprint astro-ph/9709046
\reference{}
Songaila, A., \& Cowie, L. L. 1996, AJ, 112, 335
\reference{}
Tytler, D. et al. 1995, in QSO Absorption Lines, ESO 
Astrophysics Symposia, ed. G. Meylan (Heidelberg: Springer), p.289
\reference{}
van der Marel, R. de Zeeuw, P. T., Rix, H-W., \& Quinlan, G. D.
1997, Nature, 385, 610 
\reference{}
Wright, E. L. 1981, ApJ, 250, 1
\reference{}
Wright, E. L., et al. 1994, ApJ, 420, 450
\reference{}
Yi, I. 1996, ApJ, 473, 645
\reference{}
Zaldarriaga, M., Spergel, D., \& Seljak, U. 1997, preprint  astro-ph/9702157
\reference{}
Zheng, W., Kriss, G. A., Telfer, R. C., Grimes, J. P., Davidsen, A. F. 1997,
   ApJ, 475, 469

\end{references}
\end{document}